\documentclass[seceq]{ptptex}

\usepackage{graphicx}
\title{%        %You can use \\ for explicit line-break
Interplay among Spin, Orbital and Lattice Degrees of Freedom
in $t_{2g}$ Electron Systems with Edge-Sharing Network of Octahedra%
}

\author{%       %Use \scshape  for the family name
Yukitoshi \textsc{Motome}$^1$,
Hirokazu \textsc{Tsunetsugu}$^2$,
Toshiya \textsc{Hikihara}$^3$,\\
Nic \textsc{Shannon}$^4$ and
Karlo \textsc{Penc}$^5$%
}

\inst{%         %Affiliation, neglected when [addenda] or [errata]
$^1$RIKEN (The Institute of Physical and Chemical Research), 
Wako, Saitama 351-0198, Japan\\
$^2$Yukawa Institute for Theoretical Physics, Kyoto University, 
Kyoto 606-8502, Japan\\
$^3$Division of Physics, Graduate School of Science, Hokkaido University, 
Sapporo 060-0810, Japan\\
$^4$Max-Planck-Institut f\"ur Physik komplexer Systeme, 
N\"othnitzer Str. 38, 01187 Dresden, Germany\\
$^5$Research Institute for Theoretical Solid State Physics and Optics, 
H-1525 Budapest, P.O.B. 49, Hungary
}

\abst{%         %this abstract is neglected when [addenda] or [errata]
Transition metal oxides whose lattice structure has 
edge-sharing network of octahedra 
constitute a diverse group of intriguing materials 
besides compounds with corner-sharing octahedra such as perovskites.
We present a theoretical investigation of 
the interplay among spin, orbital and lattice degrees of freedom 
in these materials. 
We focus on $t_{2g}$ electron systems 
where a keen competition among those degrees of freedom is expected to emerge 
under a relatively weak Jahn-Teller coupling. 
We study the interplay between spin and orbital degrees of freedom 
in vanadium spinels and titanium pyroxenes. 
We clarify the important role of the strong anisotropy 
in the orbital interactions due to the edge-sharing geometry. 
We also discuss the interplay between spin and lattice in chromium spinels 
focusing on the magnetization process under the external magnetic field. 
}

\begin{document}

\maketitle

\section{Introduction}

Transition metal oxides, in many cases, consist of the basic unit of
octahedron where a metal is surrounded by six oxygens. 
One of the famous families is the perovskite 
such as high-$T_{\rm C}$ cuprates and CMR manganites, 
in which octahedra form two-dimensional (2D) or three-dimensional (3D)
network by sharing oxygens at their corner. 
Another typical geometry is the edge-sharing network of octahedra. 
An old but still intriguing example is the spinel 
in which octahedra form 3D edge-sharing network. 
A 2D example is found in the sodium cobaltite 
which has a triangular lattice of Co cations. 
There, a large thermoelectric effect or superconductivity 
is recently attracting much interests. 

In this paper, we present our recent theoretical efforts 
to understand remarkable properties in several edge-sharing materials 
with focusing on a keen competition among spin, orbital and lattice degrees of freedom. 
In the octahedral coordinate, the fivefold energy levels of $d$ electrons 
of transition metals 
split into lower threefold $t_{2g}$ levels and higher twofold $e_g$ levels. 
Here, we consider the systems in which 
electrons in the $t_{2g}$ levels play a central role. 
In the $t_{2g}$ electron systems, it is known that 
the Jahn-Teller interaction is rather weak compared to the $e_g$ systems. 
Therefore, it is expected that the energy scales in 
spin, orbital and lattice degrees of freedom become closer to each other and 
that a keen competition among them seriously affects 
physical properties of the system. 

This paper is organized as follows. 
In Sec.~\ref{sec:spin-orbital}, we mainly address the interplay 
between spin and orbital degrees of freedom. 
In Sec.~\ref{sec:V-spinel}, we discuss the mechanism of lifting the degeneracy 
due to the geometrical frustration in vanadium spinels. 
We remark how the magnetic ordering pattern is changed 
depending on the sign of the third-neighbor spin coupling. 
In Sec.~\ref{sec:Ti-pyroxene}, we discuss the non spin-Peierls mechanism 
of the spin-singlet formation in titanium pyroxenes. 
We observe an interesting phase competition 
between different spin-orbital orderings. 
In Sec.~\ref{sec:Cr-spinel}, we discuss the interplay 
between spin and lattice degrees of freedom in chromium spinels 
to understand the unusual magnetization process under the external magnetic field.

\section{Interplay between spin and orbital}
\label{sec:spin-orbital}

\subsection{Vanadium spinels}
\label{sec:V-spinel}

Vanadium spinels $A$V$_2$O$_4$ with nonmagnetic $A$ cations 
such as Zn, Mg or Cd have 3D edge-sharing network of VO$_6$ octahedra, 
and consequently, magnetic V cations form 
the geometrically frustrated pyrochlore lattice. 
In general, the geometrical frustration results in 
(nearly) degenerate ground states and 
suppresses a long-range ordering. 
Nevertheless, these compounds exhibit two successive transitions 
at low temperatures: 
One is the structural transition at $T_{\rm c1} \simeq 50$K 
from high-temperature cubic phase to low-temperature tetragonal phase, and 
the other is the antiferromagnetic (AF) transition at $T_{\rm c2} \simeq 40$K. 
\cite{Ueda1997}
The issue is the microscopic mechanism of these two transitions, that is, 
how the frustration is reduced and 
how the long-range orders are stabilized. 

We have studied this problem by taking into account 
the orbital degree of freedom of $t_{2g}$ electrons, 
because V$^{3+}$ cation has two $d$ electrons in threefold $t_{2g}$ levels. 
\cite{Tsunetsugu2003,Motome2004} 
The important point is the spatial anisotropy of the $t_{2g}$ orbitals. 
In the edge-sharing configuration of VO$_6$ octahedra, 
the most relevant contribution in the hopping integrals 
is given by the overlap between the same orbitals lying in the same plane, 
that is, the so-called $\sigma$ bond. 
We derived an effective spin-orbital coupled model 
in the strong correlation limit 
with considering only the $\sigma$-bond contribution 
in the perturbation for the multiorbital Hubbard model. 
We found that the effective model shows the strong anisotropy 
(three-state Potts type) in the orbital intersite interactions. 
This strong anisotropy plays a crucial role 
in the keen competition between spin and orbital, and 
finally reduces the frustration. 
That is, the degeneracy is lifted in the orbital sector first,
with accompanied by the tetragonal lattice distortion. 
This explains well the structural transition at $T_{\rm c1}$ in experiments. 
\cite{Motome2004}

The mechanism of the magnetic transition at $T_{\rm c2}$ is also interesting. 
In the orbital ordered state, $d_{xy}$ orbital is singly occupied 
at every V site. 
On the other hand, 
$d_{yz}$ and $d_{zx}$ orbitals are occupied in the staggered way 
in the $z$ direction. 
The uniform occupation of $d_{xy}$ orbitals leads to 
a large enhancement of the AF spin correlation 
along one-dimensional (1D) chains lying in the $xy$ planes. 
Thus, the magnetic frustration is partially lifted by the orbital ordering. 
However, the relative angles of the staggered moments 
between different $xy$ chains 
are not yet determined because of the pyrochlore structure. 
We proposed that the third-neighbor spin exchange $J_3'$ as well as 
thermal and/or quantum fluctuations can lift the remaining degeneracy 
and establish the 3D magnetic long-range order. 
\cite{Tsunetsugu2003,Motome2004}

\begin{figure}
\centerline{\includegraphics[width=120mm]{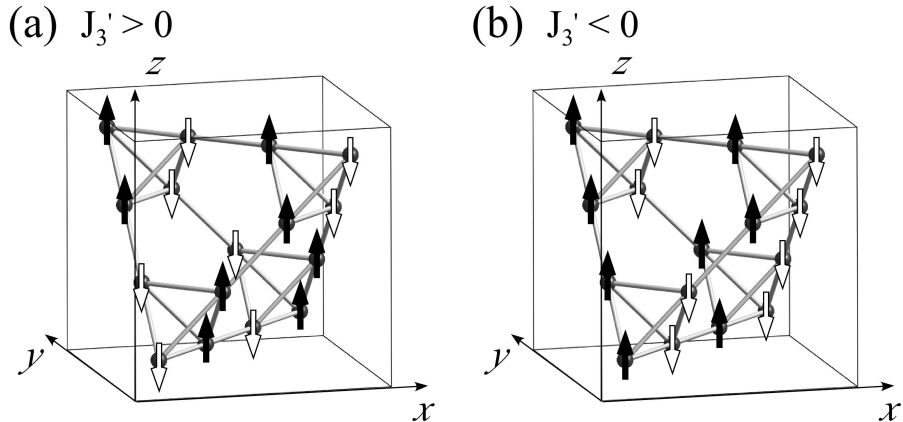}}
\caption{
Magnetic ordering structure predicted in the effective 
spin-orbital coupled model for vanadium spinels with 
(a) antiferromagnetic $J_3'$ and (b) ferromagnetic $J_3'$.
}
\label{fig:1}
\end{figure}

The 3D magnetic ordering pattern is hence determined 
by the way of stacking the 1D AF $xy$ chains in the $z$ direction, 
and strongly depends on the sign of $J_3'$. 
When $J_3'$ is AF as in our effective model, 
the spin configuration is up-up-down-down-... 
(four-times period) along the $yz$ and $zx$ chains 
as shown in Fig.~\ref{fig:1} (a).  
This pattern is consistent with the neutron scattering results 
by Niziol 
\cite{Niziol1973} 
and Lee {\it et al}. 
\cite{Lee2004}
On the other hand, if we suppose the ferromagnetic $J_3'$, 
the primitive unit cell (four-site tetrahedra) corresponds to 
the magnetic unit cell, and 
hence a different spin configuration may be obtained;  
up-down-up-down-... in the $yz$ (or $zx$) chains and ferromagnetic in the others 
as shown in Fig.~\ref{fig:1} (b). 
However, we note that 
it is rather difficult to obtain the ferromagnetic $J_3'$ 
by the perturbation for the multiorbital Hubbard model and
that the latter spin ordering may lead to further lowering of 
lattice symmetry through the spin-lattice coupling. 
Thus, theoretical predictions depend on 
the sign of the small energy scale $J_{3}'$, 
but the spin configuration in Fig.~\ref{fig:1} (a) is the most likely 
within our theoretical model. 
The point to be stressed is that the small coupling $J_3'$ 
becomes crucial only when the frustration is reduced by the orbital ordering. 

\subsection{Titanium pyroxenes}
\label{sec:Ti-pyroxene}

Titanium pyroxenes $A$TiSi$_2$O$_6$ ($A$=Na or Li) are 
another typical compounds with edge-sharing octahedra 
where the orbital degree of freedom may play a crucial role. 
These materials have quasi 1D structure in which  
the octahedra share their edges alternatively to form zig-zag chains. 
Due to this peculiar structure, threefold $t_{2g}$ levels split into 
a lower doublet and a higher singlet. 
Ti$^{3+}$ cation has one $d$ electron in the lower twofold levels. 
The compounds show a phase transition at $T_{\rm c} \simeq 200$K 
where the magnetic susceptibility suddenly drops and 
the lattice structure is dimerized along the 1D zig-zag chains. 
\cite{Isobe2002,Ninomiya2003}
The temperature dependence of the magnetic susceptibility 
as well as the estimate of the spin gap compared to $T_{\rm c}$ suggests that 
the transition cannot be understood by the usual spin-Peierls mechanism. 
\cite{Isobe2002}

We have considered the mechanism of this peculiar transition 
by taking account of the twofold orbital degeneracy explicitly. 
As in the case of V spinels, 
the spatial anisotropy of the $t_{2g}$ orbitals and 
the edge-sharing geometry play an important role, 
resulting the Ising-type orbital interaction 
in the effective spin-orbital model. 
We found that the model exhibits two different ground states; 
one is the spin-dimer and orbital-ferro state and 
the other is the spin-ferro and orbital-antiferro state. 
The transition between them can be controlled by 
the Hund's-rule coupling $J_{\rm H}$ and/or the external magnetic field $h$. 
For the realistic values of parameters, 
the ground state is the former one. 
The obtained temperature dependence of the magnetic susceptibility explains 
the experimental result semiquantitatively. 
\cite{Hikihara2004}

\begin{figure}
\centerline{\includegraphics[width=110mm]{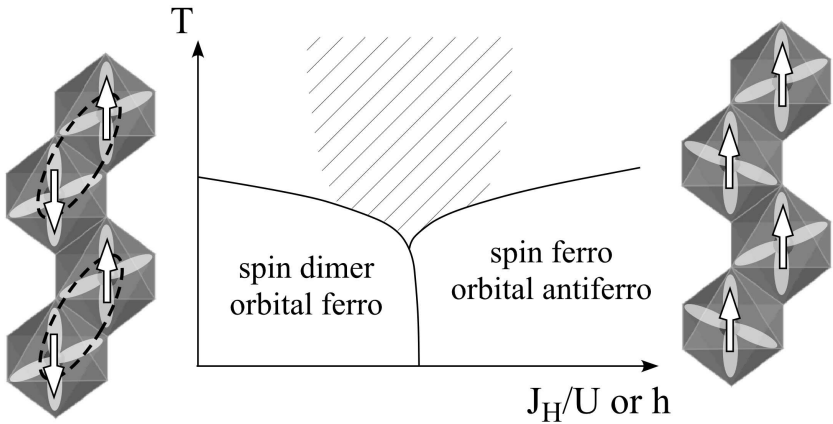}}
\caption{
Schematic multicritical phase diagram of the effective 
spin-orbital model for titanium pyroxenes. 
The hatched area shows the multicritical competing regime. 
}
\label{fig:2}
\end{figure}

Let us focus on the spin-orbital phase competition 
in the parameter space of $J_{\rm H}$ and/or $h$. 
A schematic phase diagram is shown in Fig.~\ref{fig:2}. 
In the multicritical regime 
(the hatched area in Fig.~\ref{fig:2}), 
the competition becomes conspicuous; 
both spin and orbital correlations are suppressed as the temperature decreases. 
This severe competition suppresses $T_{\rm c}$ 
in the multicritical regime, and 
enables us to see effects of the competition 
more clearly above $T_{\rm c}$. 
The details will be reported elsewhere. 
\cite{HikiharaPREPRINT} 
In order to observe the intrinsic effects of the spin-orbital competition, 
it is desired to realize an experimental situation 
in the vicinity of the multicritical point. 

\section{Interplay between spin and lattice: Chromium spinels}
\label{sec:Cr-spinel}

Chromium spinels $A$Cr$_2$O$_4$ with nonmagnetic $A$ cations 
show only one phase transition 
\cite{Kino1971}
in contrast to the two transitions in V spinels in Sec.~\ref{sec:V-spinel} 
nevertheless the lattice structures are isomorphic. 
The difference is in the orbital state: 
Because Cr$^{3+}$ has three $d$ electrons in threefold $t_{2g}$ orbitals, 
the orbital degree of freedom is inactive in Cr spinels. 
At the transition temperature, an AF spin ordering occurs simultaneously 
with the structural change from cubic to tetragonal. 
The origin of this transition has been discussed in terms of 
the spin Jahn-Teller mechanism; 
the geometrical frustration is reduced 
by the lattice distortion which gains the magnetic exchange energy. 
\cite{Yamashita2000,Tchernyshyov2002}

Very recently, the compounds with $A$=Hg or Cd are found to exhibit 
the half-magnetization plateau, which is unusually stable, 
under the external magnetic field. 
\cite{Ueda2005,UedaUNPUBLISHED}
This plateau phenomenon has been also discussed 
by using the AF Heisenberg model with the spin-lattice coupling. 
\cite{Penc2004}
The plateau is well reproduced by the mean-field results 
in the ground state for finite spin-lattice couplings. 

Following this idea, we have extensively studied 
finite-temperature properties of 
the spin-lattice coupled model on the pyrochlore lattice 
under the magnetic field 
by using the Monte Carlo simulation. 
We found that the half-magnetization plateau remains robust 
at finite temperatures, and that 
the temperature dependence of the magnetization curve is 
favorably compared with the experimental results. 
\cite{MotomePREPRINT}
We also found that the plateau phase is induced 
by thermal fluctuations even in the absence of the spin-lattice coupling. 
The details of the results and the comparison with the experimental results 
will be reported elsewhere. 
\cite{MotomePREPARATION}

\section*{Acknowledgements}
Y.M. would like to thank D.I. Khomskii and S.-H. Lee for stimulating discussions. 
This work is supported by a Grant-in-Aid and NAREGI 
from the Ministry of Education, Science, Sports, and Culture of Japan, and
the Hungarian OTKA Grant No. T038162 and the JSPS-HAS joint project.

%\appendix
%\section{First Appendix} %Empty argument \section{} yields `Appendix'. 
%
%\section{Second Appendix}

\end{document}